\documentclass[a4paper,11pt]{article}
\pdfoutput=1 

\usepackage{jinstpub} 
\usepackage{siunitx}
\DeclareSIUnit\bit{bit}

\usepackage{hyperref}

\usepackage{lineno}

\title{Timing and Synchronization of \\ the DUNE Neutrino Detector}

\author[a]{A.Barcock,}
\author[b,1]{D.Cussans,\note{Corresponding author.}}
\author[b]{ D. Lindebaum,}
\author[a]{D. Newbold,}
\author[a]{S.Paramesvaran,}
\author[a]{and  S.Trilov}

\affiliation[a]{STFC RAL, Didcot, UK}
\affiliation[b]{University of Bristol, UK}

\emailAdd{david.cussans@bristol.ac.uk}

\abstract{The DUNE neutrino experiment far detector has a fiducial
  mass of \SI{40}{\kilo \tonne}.  The O(1M) readout channels are
  distributed over the four \SI{10}{\kilo \tonne} modules and need to
  be synchronized with respect to each other to a precision of
  O(\SI{10}{\nano \second}). The entire system needs to be
  synchronized with respect to GPS time to O(\SI{100}{\nano
    \second}). The system needs to be reliable, simple and affordable.
  Clock and synchronization information are encoded on the same fibre
  using a protocol based on duty cycle shift keying (DCSK) with 8b10b
  encoding to ensure DC-balance. The use of DCSK allows the clock to
  be recovered directly by PLL based clock generators without needing
  to use a separate clock and data recovery (CDR) device. Small scale
  tests show a timing jitter at the endpoint of $\approx$~\SI{10}{\pico \second} with respect to the timing master.
}

\notoc

\begin{document}
\maketitle
\flushbottom

\section{Introduction}

The Deep Underground Neutrino Experiment
(DUNE)~\cite{ref:DUNE_TDR_Vol1,ref:DUNE_TDR_Vol4} will detect neutrinos
generated \SI{1300}{\kilo \metre} away in Fermilab, near Chicago. The
far detector will consist of four modules, each with \SI{10}{\kilo
  \tonne} fiducial mass of liquid Argon, located \SI{1.5}{\kilo
  \metre} underground at the Sanford Underground Research Facility
(SURF).

The energy deposited by the products of neutrino interactions will be
detected in two ways: ionization drifted by an applied electric field
to pick-up electrodes, forming a time projection chamber (TPC), and
scintillation light.

In order to reconstruct events with the photon detection system
readout channels must be synchronized to each other with a precision
of nanoseconds. To correlate events in the far detector with neutrino
generation at Fermilab the entire detector must be synchronized to GPS
time to a precision of O(\SI{100}{\nano \second}).

\section{The DUNE Timing System}

The DUNE timing system is made of a mix of commercially available and
custom components.  It was not possible to use a COTS solution due to the requirement during prototyping to distribute a low, and fixed, latency ``trigger'' to all endpoints using the timing system. However, commercial components are used where possible. 

The DUNE timing system protocol is designed to allow use with passive
optical splitting and combining. This will reduce system cost and allow hot-swap
redundancy. The two completely independent master timing
systems will cross check each other and  allow firmware and software updates without interrupting
system running. Two independent antennae will be used: One at the top of
each of the two access shafts to SURF. This, together with a rubidium
atomic clock in the master timing system will enable failure of the
link to GPS time to be detected. 

\begin{figure*}
\includegraphics[width=\textwidth]{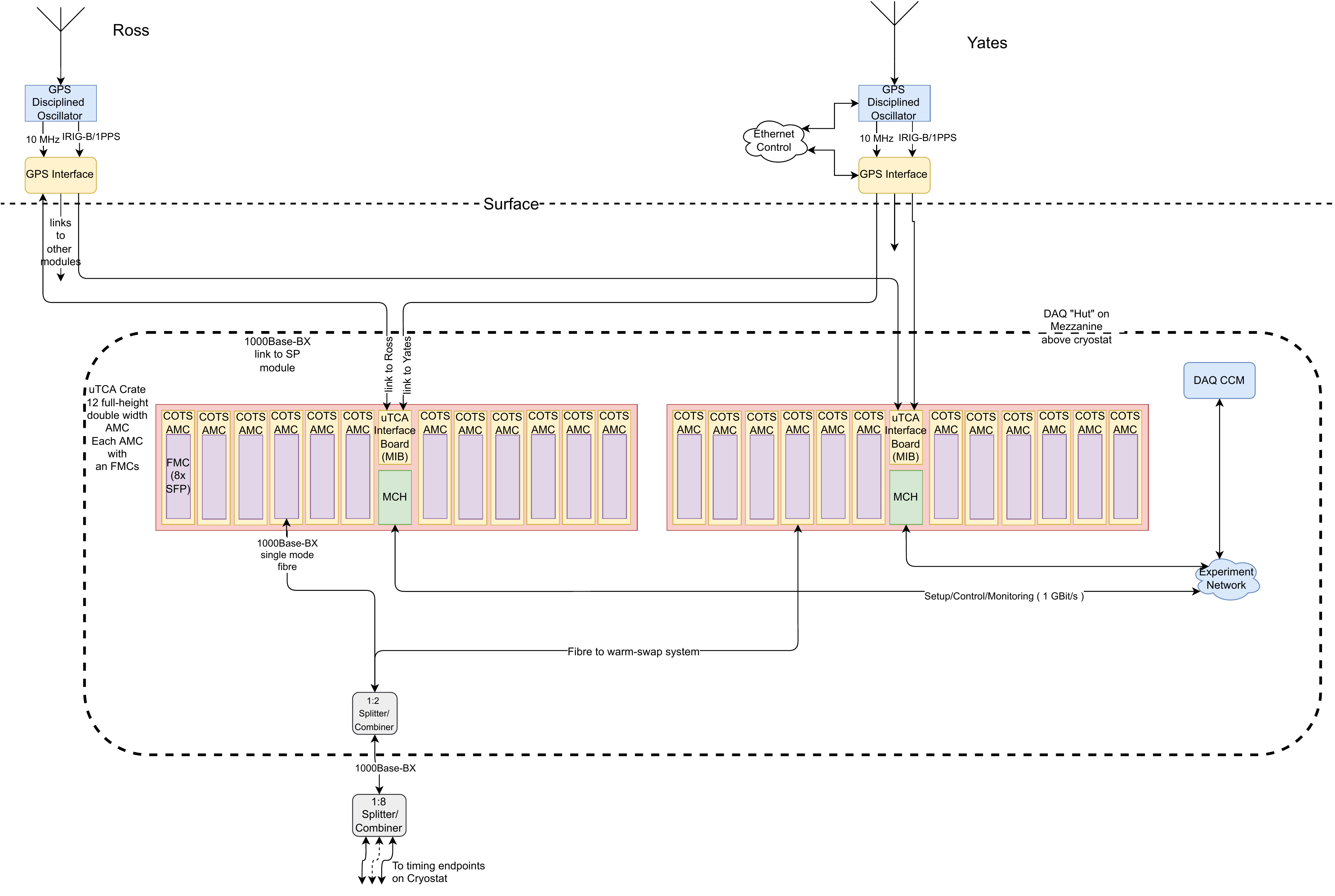}%
	\caption{Schematic view of the DUNE timing system.}
	\label{fig:DTS_overall}
\end{figure*}

\begin{figure*}
\includegraphics[width=\textwidth]{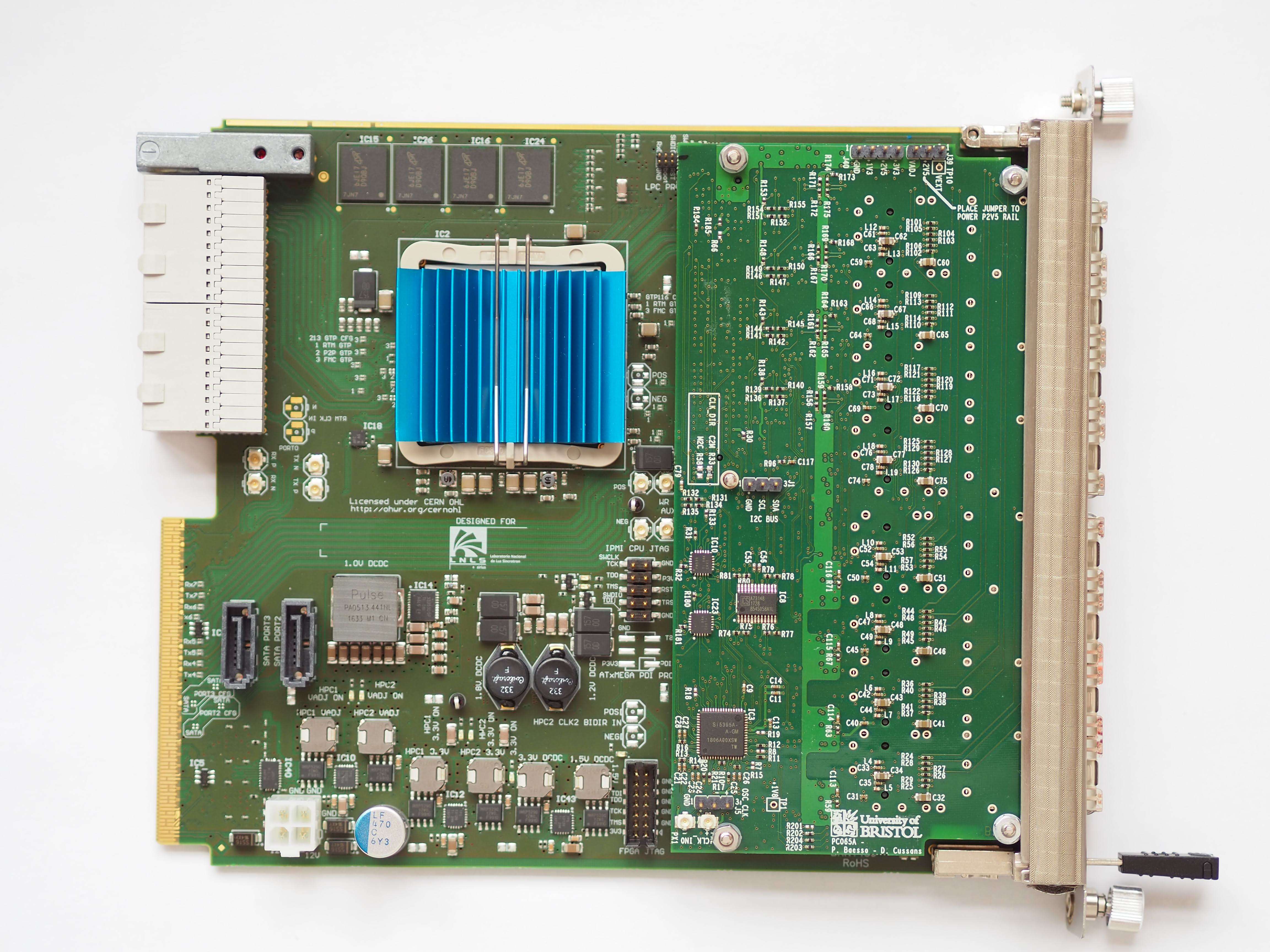}%
	\caption{Photograph of a DUNE timing system Fibre Interface Board mounted on a commercial FPGA based AMC.}
	\label{photo:afc_fib}
\end{figure*}

At the top of each access shaft a GPS-disciplined
oscillator provides timing information to a custom GPS Interface
Module (GIB). Fibres from each GIB are connected to MicroTCA Interface
Boards (MIBs), a custom AMC\cite{ref:utca-amc}. They sit in one of the two crate
controller slots of MicroTCA\cite{ref:utca-spec} crates located on each detector module.
Timing signals are distributed over the backplane to AFC boards~\cite{ref:afc} (a commercial AMC). Each AFC carries a fibre interface board
(FIB), a custom FMC\cite{ref:fmc-spec}. Each FIB has eight fibres, transmitting timing
information to the detector readout electronics.  Figure
\ref{fig:DTS_overall} illustrates the overall layout of the DUNE
timing system. Photograph~\ref{photo:afc_fib} shows a FIB mounted onto
an AFC. The use of a COTS carrier board allows the FIB to be simple, consisting mainly of a clock fan-out and a low jitter (\SI{0.8}{\pico \second} RMS added jitter) D-type flip-flop to re-time data from an FPGA on the carrier board using a low jitter clock distributed over the MicroTCA backplane.

The distance from the timing master to the timing endpoints is in the region of \SI{200}{\metre} and the optical fibres will vary by up to \SI{100}{\metre}. Delays in the timing endpoints are used to ensure synchronization of the clock and time stamps at every point in the detector.

\section{Protocol}

\subsection{Duty Cycle Shift Keying}

\begin{figure}
    \centering
    \includegraphics[width=\textwidth]{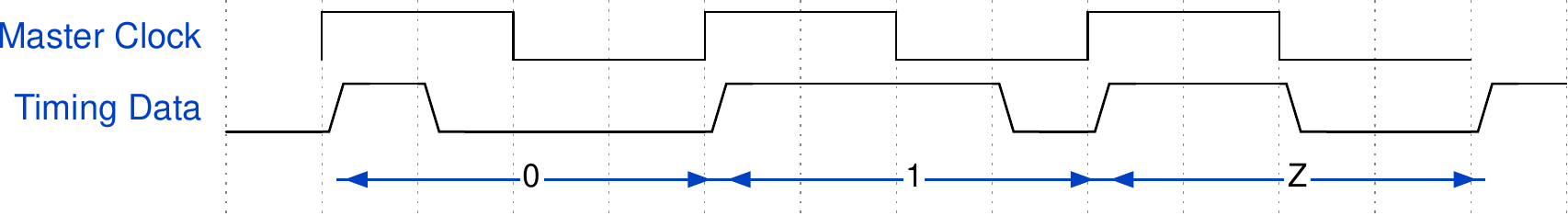}
    \caption{Illustration of Duty Cycle Shift Keying (DCSK) showing the representation of '0' , '1' and 'Z' values.}
    \label{fig:dcsk_01z}
\end{figure}

\begin{figure}
    \centering
    \includegraphics[width=\textwidth]{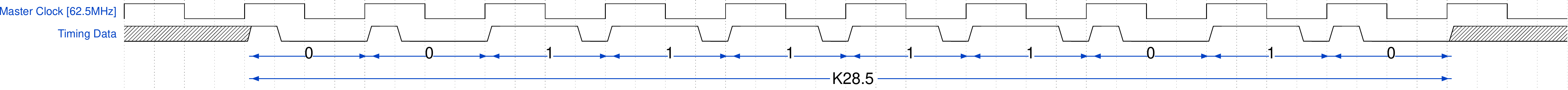}
    \caption{Example of 8b10b Encoded DCSK Data, showing a K28.5 control character used for link alignment}
    \label{fig:8b10b_encoded_dcsk}
\end{figure}

Earlier prototypes of the DUNE timing system~\cite{CUSSANS2020162143}
used non-return-to-zero encoding and ADN2814 CDR
ASICs~\cite{ref:adn2814} to recover the clock. However, due to a
potentially long term shortage of the ADN2814, the system was
redesigned to use Duty Cycle Shift Keying (DCSK)~\cite{ref:dcsk} which
allows the clock to be recovered directly by a PLL based clock
generator. It is also possible to recover the clock using the clock
generation modules inside Xilinx FPGAs. Data are transmitted at
\SI{62.5}{\mega \bit / \second} with ``0'' represented by a $25\%$
duty cycle and ``1'' by a $75\%$ duty cycle.  8b/10b~\cite{ref:8b10b}
encoding is used to ensure DC-balance. Figure \ref{fig:dcsk_01z}
illustrates the values ``0'' , ``1'' and ``Z'' encoded with
DCSK. In this way both clock and data are distributed down the same link. Figure~\ref{fig:8b10b_encoded_dcsk} illustrates an 8b/10b
control character using DCSK.

\subsection{Higher Level Protocol}

A continuous stream of encoded serial data is transmitted over
1000Base-Bx~\cite{ref:ethernet} optical links from the timing master to
timing “end points”. The data carries commands that are used to
synchronize a 64-bit time-stamp counter in the end-point and propagate
commands and synchronization messages. This allows the readout clocks,
calibration signals, etc. to be synchronized throughout each module.

Each link from the timing master is bidirectional with data
transmitted in both directions down a single fibre with different
light wavelengths. Normally, data are only transmitted from the timing
master to the endpoints. However, the timing master can command an endpoint to transmit data back to the master. This allows the
timing master to measure the latency between timing master and endpoint and adjust the clock phase and time-stamp at the endpoint. This
allows all endpoints to operate with the same time-stamp and clock
phase. The return path also allows the timing master to read out the status of the end points. To reduce system cost each fibre is passively split into up to eight fibres. Each of the fibres coming from an optical splitter is connected to a single timing endpoint. The master transmits timing information continuously and the higher level protocol ensures that only one endpoint at a time transmits back to the master on a fibre.  

\section{Laboratory Tests}

\begin{figure*}
	\includegraphics[width=0.9\columnwidth]{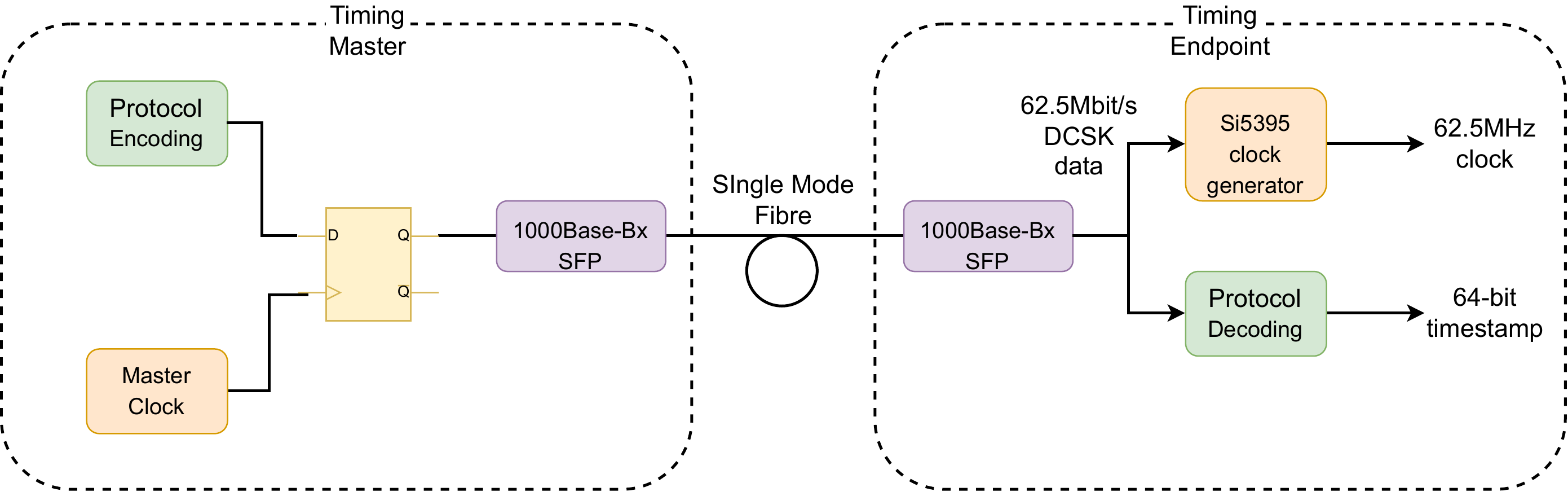}
	\caption{Block diagram of jitter performance test setup.}
	\label{fig:jitter_setup}%
\end{figure*}

\begin{figure}
	\includegraphics[width=0.9\columnwidth]{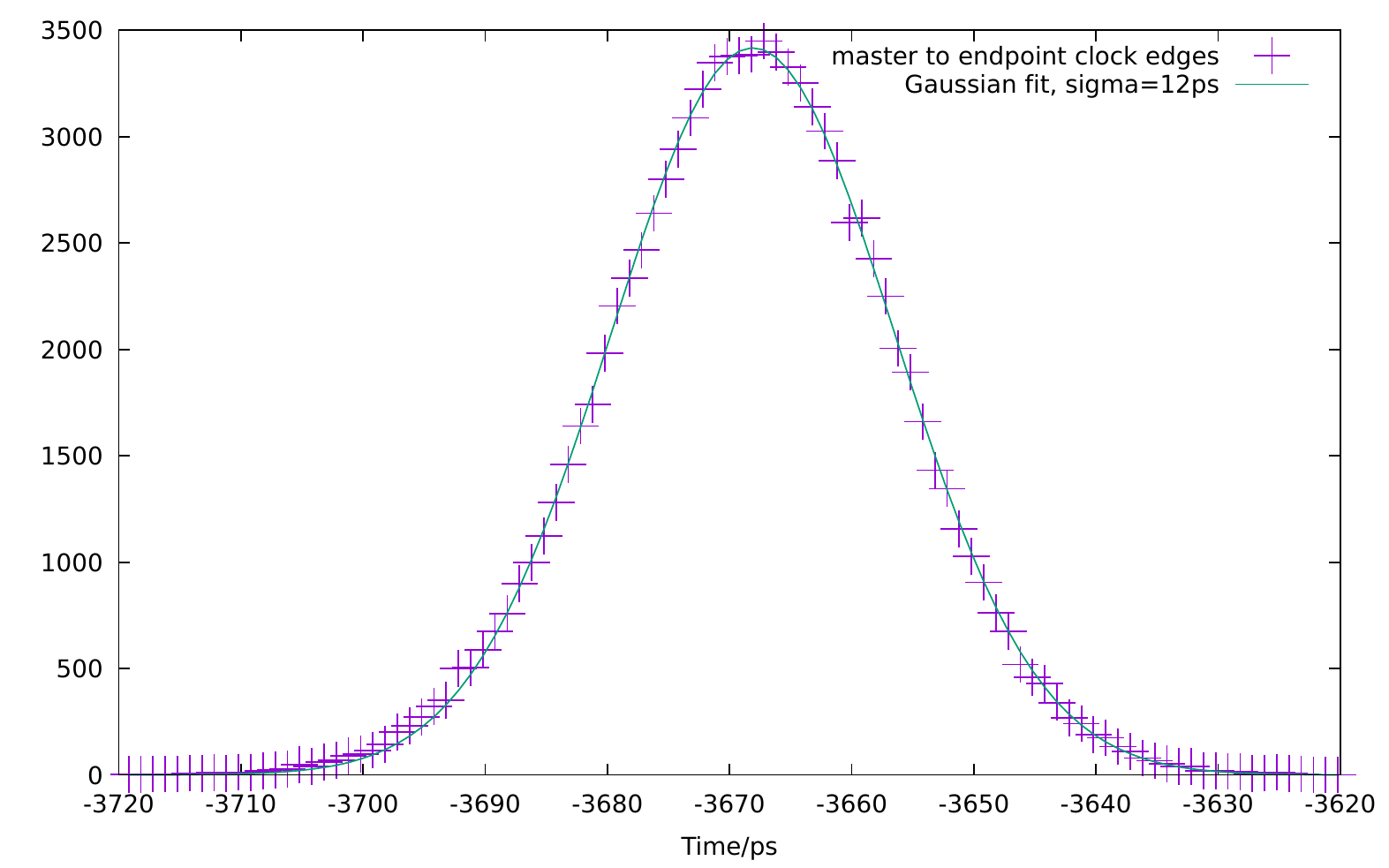}%
	\caption{Distribution of time difference between clock edges at timing master and endpoint.}
	\label{fig:timing_jitter}
\end{figure}

Timing jitter performance was measured in the laboratory by connecting
a prototype timing master with a prototype endpoint by optical
fibre. The clock signals in the master and endpoint were
compared. Timing jitter depends on the bandwidth programmed into the
PLL clock generators used but is typically in the region of \SI{10}{\pico
  \second}. A block
diagram of the apparatus used to test cycle-to-cycle timing jitter is
shown in figure~\ref{fig:jitter_setup}. The distribution of the time
difference between clock edges at the timing master and timing
endpoint is shown in figure~\ref{fig:timing_jitter}. In other tests
over 100 endpoints were synchronized.

\section{Summary}

The DUNE timing system will provide nano-second precision
synchronization of several hundred readout units distributed over a
physically large detector. Laboratory tests indicate that the system
will meet its specification comfortably. The novel use of Duty Cycle
Shift Keying allows a clock to be recovered from a serial data stream
without the use of a specialized clock and data recovery device. The
use of passive optical splitting allows each fibre from the timing
system to drive multiple endpoints while simultaneously allowing for
a redundant system.  Two GPS disciplined oscillators will be continuously compared in order to improve system reliability.


\bibliography{Timing_and_Synchronization_of_the_DUNE_Neutrino_Detector_cussans_twepp2022}
\bibliographystyle{JHEP}

\end{document}